# Spin-current amplification by stimulated spin pumping

Benjamin Assouline, Marina Brik, Nirel Bernstein, Amir Capua*

Department of Applied Physics, The Hebrew University of Jerusalem, Jerusalem 91904, Israel

*e-mail: amir.capua@mail.huji.ac.il

**Abstract:**

**Stimulated emission is a process during which an atomic system gives away energy to create a coherent photon. It is fundamental to the operation of the optical amplifier. Here we propose two mechanisms for amplifying AC spin currents in a solid-state magnetic medium by a stimulated spin pumping process. The first is synchronous and consists of phase-locked pulses that perturb a precessing magnetic moment. The second is asynchronous and is driven by DC spin currents. The amplification relies on a non-adiabatic interaction taking place in a ferromagnetic medium in which the magnetic moment emits spin angular momentum in the form of spin current before equilibrating with the environment. The pumped spin current amplifies or absorbs the injected AC spin current mimicking the operation of the optical gain medium as readily seen from the gain saturation profiles. The mechanisms we propose are a first step towards a realistic spin current amplifier.**

**Introduction**

When an excited two-level system (TLS) is subject to the action of a photon, an additional photon may be emitted in a process known as stimulated emission. This is the principle that stands behind optical amplification and is a non-adiabatic (NA) process [1]. Namely, it is a process whereby energy is transferred between an atomic system and a driving electromagnetic (EM) radiation before the two equilibrate. In a variety of applications, the NA interaction is pivotal, e.g. in the coherent manipulation of the qubit and its gate operations [2,3] or in coherent control procedures in quantum optics [4-6]. In the context of magnetic materials and spintronics technology, the NA interaction is less common.

The NA interaction can be considered as a hybrid time-frequency process. Traditionally, magnetization dynamics have been explored either in the time domain or the frequency domain and unraveled a variety of useful quantum effects for spintronics technology. The frequency domain was dominated by the ferromagnetic resonance (FMR) experiment [7-10] and its variations [11-13], in which the EM field drives the magnetic system in steady state and energetic equilibrium pertains. Therefore, the FMR experiment probes the adiabatic process. Likewise, the time domain pump-probe free induction decay experiments [14-20] are also not NA because the EM field that drives the system is absent. For these reasons, a pure time or frequency domain experiment cannot excite the NA interaction but rather the hybrid experiment is required [21,22].

Recently, interest has grown in the NA interaction in magnetic solid-state media. At the quantum limit of the single magnetic electron spin, coherent control was demonstrated using a scanning tunneling microscope [23,24] and was identified useful for quantum information processing. On the macroscopic scale, a spatial non-equilibrium energy exchange was demonstrated between counter propagating spin waves in yttrium iron garnets using artificial magnonic crystals [25] and was discussed in the context of signal processing. In Ref. [21] it was shown that the NA regime can be excited in ferromagnets (FMs) by combining the FMR with the time-resolved magneto optical Kerr effect. In these experiments the response was examined before the steady precessional magnetization state was reached and a variety of phenomena were observed that are more familiar in their optical form. Such phenomena included a controllable frequency chirping of the magnetization, induction of coherence in the inhomogeneously broadened spin ensemble, tuning of the intrinsic relaxation times, and mode-locking of the spin ensemble.

These advances paved the way for applying concepts from the fields of quantum optics and information processing to spintronics technology. Recently, the EM radiation of the NA interaction was replaced by a current of polarized spins and a coherent spin-transfer-torque (STT) control scheme in Py was demonstrated without the complications of driving the EM radiation [26]. The work was complemented by a rigorous theory of the NA regime in FM systems that are driven by EM radiation and spin currents [22].

Here we discuss an additional application of the NA interaction in spintronic systems and propose an AC spin current amplifier where the spin current is amplified by stimulated spin-pumping. Our work is inspired by the operation of the optical amplifier, such as the erbium doped fiber amplifier (EDFA) or the semiconductor optical amplifier (SOA) where the excited carriers amplify the EM radiation by stimulated emission. We propose two mechanisms for amplifying AC spin currents in FMs. The first mechanism is synchronous and utilizes optical pumping pulses that are phase-locked with the injected AC spin current and excite the NA regime [14,21,27]. By emission of spin current from the FM and precise timing of the optical pulses, the AC spin current is either amplified or absorbed. The second mechanism relies on exciting first the STT oscillator by a DC spin Hall effect (SHE). It is asynchronous since it does not require prior preparation of the phase of the excitation. Consequently, the injected AC spin current is amplified or absorbed depending solely on the DC excitation level. We show that the operation of the asynchronous spin current amplifier is reminiscent of the operation of the optical amplifier and that the similarity to the optical amplifier extends to the gain saturation profiles. We use a closed-form analytical model that is based on conversion of the optical TLS Bloch formalism to its spintronic analogue, which further supports these conclusions.

**Results**

At the core of the spin current amplifier is a bilayer of a normal-metal (NM)-FM as presented in Fig. 1 that serves both mechanisms. The NM is chosen to have a large SHE. An AC charge current of amplitude $J_{C\,AC}$ is passed through the NM and generates an AC spin current of amplitude $J_{S\,AC}$ by the SHE. The AC spin current is polarized in the $\hat{x}$ direction and drives the FMR in the presence of an effective magnetic field, $\vec{H}_{eff}$. The anisotropy fields are neglected so that $\vec{H}_{eff}$ is equal to the external magnetic field that is applied in the $\hat{z}$ direction and defines the energy axis. The magnitude of the external magnetic field is $H_0$ and is set to resonance. At the NM-FM interface, spin pumping takes place. So far this process is identical to the STT-driven FMR [12]. Coherent spin current amplification takes place when the NA interaction is excited such that the pumped spin current, $\vec{J}_{SP}$, adds constructively to $\vec{J}_{S\,AC}$ (vectors indicate spin orientation).

We start by addressing the synchronous amplification scheme. In this scheme the NA regime in the FM is excited by applying a train of ultrashort optical pulses that perturb the steady precessional state [14,16,21] as illustrated in Fig. 1. The pulses are phased-locked with $\vec{J}_{S\,AC}$ while each pulse is modeled as an effective instantaneous torque in the $\hat{y}$ direction. Using the Landau–Lifshitz–Gilbert–Slonczewski (LLGS) formalism under the macrospin approximation, we simulated 100 fs pulses at a repetition rate, $\Gamma_{rep}$, of 100 MHz that represent realistic experimental conditions [26]. The temporal evolution of the $\hat{z}$ component of $\vec{M}$, $M_z$, is depicted in Fig. 2(a) & 2(b) for absorption and amplification. Fig. 2(a) shows the case where the relative phase, $\phi_{RF}$, between the optical pulses and $\vec{J}_{S\,AC}$ is such that $\vec{J}_{SP}$ is in-phase with $\vec{J}_{S\,AC}$ and amplification is maximal. We denote this condition by $\phi_{RF} = 0°$ as illustrated in Fig. 2(c). Before turning on $J_{C\,AC}$, $\vec{M}$ is aligned with $\vec{H}_{eff}$. Once $J_{C\,AC}$ is turned on, the generated $\vec{J}_{S\,AC}$ drives the steady precessional state by STT. At t = 150 ns the optical pulses are applied and initiate the spin mode-locking [21] that excite the NA interaction. Over time, the steady NA interaction builds up and stabilizes at ~ 230 ns as seen in the inset. In between pulses, $M_z$ rises and the Zeeman energy decreases. This energy is transferred to $\vec{J}_{S\,AC}$ by the emission of spin pumping current, $\vec{J}_{SP}$.

Depending on the relative phase between $\vec{J}_{SP}$ and $\vec{J}_{S\,AC}$ which is denoted by $\phi_{J_{SP}-J_{S\,AC}}$, $\vec{J}_{S\,AC}$ is either amplified or absorbed.

The spin pumping is given by $\vec{J}_{SP} = \frac{\hbar}{4\pi}\left\{\frac{1}{M_s^2}\Re(g_{SP}^{\uparrow\downarrow})\vec{M}\times\frac{d\vec{M}}{dt} - \frac{1}{M_s}\Im(g_{SP}^{\uparrow\downarrow})\frac{d\vec{M}}{dt}\right\}$ [28] where $M_s$ is the magnetization saturation and $g_{SP}^{\uparrow\downarrow}$ is the spin pumping conductance at the NM-FM interface. The first term is the dominant one and describes a damping-like torque while the second term has a minor effect on the gyromagnetic ratio, $\gamma$, and may be neglected [28]. Therefore, the primary effect of $\vec{J}_{SP}$ is to increase the intrinsic Gilbert damping, $\alpha$, by $\alpha_{SP} = \Re(g_{SP}^{\uparrow\downarrow})\frac{\gamma\hbar}{4\pi t_{FM}M_s}$.

The parameters of the simulation were chosen to model a realistic NM-FM bilayer sample such as the Pt/Py system [29] with $\alpha = 0.0075$, thickness of the FM layer of $t_{FM} = 12$ Å, saturation magnetization of $M_s = 3\cdot 10^5 A/m$, spin pumping conductance of $\Re(g_{SP}^{\uparrow\downarrow}) = 10^{19} m^{-2}$, and spin Hall angle of $\Theta_{SHE} = 0.15$. Moderate AC and DC charge current densities of $10^7 - 10^{10}\,A/m^2$ were used. The external field, $\mu_0 H_0$, was 30 $mT$ which corresponds to a resonance frequency of 1 GHz with $\mu_0$ being the permeability of the vacuum. These parameters were chosen such that the relaxation time is long enough and the response builds up from pulse to pulse as shown in the inset of Fig. 2(a). This condition is met when $\gamma\mu_0 H_0\alpha < \Gamma_{rep}$. On the other hand, when this condition is not fulfilled, each perturbation acts as an isolated event and the overall amplification drops significantly. Fig. 2(b) presents the results for the case of maximal absorption conditions which is achieved by setting $\phi_{RF}$ to 180°. The same principles described for amplification hold for absorption as well. In a broader context, the synchronous spin current amplification/absorption process is based on the same timing principles that dominate the Ramsey interferometry method of separated phase-shifted fields [5,30].

The gain characteristics of the synchronous amplifier are presented in Fig. 3. The amplified spin current $\vec{J}_{S\,out}$ is a sum of $\vec{J}_{S\,AC}$ and $\vec{J}_{SP}$. In our case $\vec{J}_{S\,AC}$ is polarized in the $\hat{x}$ direction. Accordingly, the spin current gain is given by $(\vec{J}_{S\,AC} + \vec{J}_{SP})_x / (\vec{J}_{S\,AC})_x$. Figure 3(a) illustrates the dependence of the gain on the relative timing of the pulses. The Figure presents the data for optically induced magnetic pulses of 0.01 T - 0.05 T in addition to a strong pulse of 0.15 T. For weak pulses (0.01 T and 0.02 T) the torque exerted by the pulses is insufficient to reach amplification independently of $\phi_{RF}$ and the magnetization dynamics are primarily dictated by the STT induced by $\vec{J}_{S\,AC}$. This is readily seen from the plots of $\phi_{J_{SP}-J_{S\,AC}}$ of Fig. 3(b). For the 0.01 T and 0.02 T cases, $\phi_{J_{SP}-J_{S\,AC}}$ is always between 90° and 180° so that only destructive interference of $\vec{J}_{SP}$ and $\vec{J}_{S\,AC}$ takes place.

As the pulses' magnitude increase their influence becomes noticeable. By a process referred to as injection locking [31], $\vec{M}$ starts to follow the pulses and $\phi_{J_{SP}-J_{S\,AC}}$ is set such that amplification can be reached as seen for 0.03 T and 0.05 T. As the intensity of the pulses is further increased, the STT arising from $\vec{J}_{S\,AC}$ becomes negligible as compared to the torque induced optically and $\vec{M}$ is eventually locked solely to the pulses. This limit is illustrated in Fig. 3(b) for the case of strong 0.15 T pulses for which $\phi_{J_{SP}-J_{S\,AC}}$ depends linearly on $\phi_{RF}$ and the gain is determined only by the timing of the pulses. Generally, $\vec{J}_{SP}$ and $\vec{J}_{S\,AC}$ add constructively when the pulses drive $\vec{M}$ such that the transverse components of $\vec{M}$ and $\vec{J}_{SP}$ (i.e. $\hat{x}, \hat{y}$) change sign as compared to the unperturbed case. In the same manner, when the pulses are turned off at $t = 640\,ns$, $\vec{M}$ chirps and $\vec{J}_{SP}$ and $\vec{J}_{S\,AC}$ add destructively again. This process is illustrated in Fig. 3(c) which is a close up of $t = 645 - 655\,ns$ of Fig. 2(a). $M_z$ increases to its maximal value at $t = 650\,ns$ after which the phase of $M_x$ is inverted and absorption takes place. A close examination of $M_x$ reveals a frequency chirping that is typical of a NA interaction [5] and has been addressed in FMs in Ref. [21].

The gain saturation profiles of the synchronous mechanism are presented in Fig. 3(d). The traces are plotted as a function $J_{C\,AC}$. In contrast to the gain saturation profiles of the optical amplifier illustrated qualitatively in the inset, the spin current gain saturates monotonically already for very low $J_{C\,AC}$ amplitudes. While in the optical amplifier such as the EDFA or SOA the gain saturates due to carrier depletion at optical powers above the saturation power, the synchronous amplifier saturates already at weak inputs because the gain relies purely on the phase between the perturbed $\vec{M}$ and $\vec{J}_{S\,AC}$. As $J_{C\,AC}$ is increased, the AC STT overcomes the optically induced torque so that $\phi_{J_{SP}-J_{S\,AC}}$ becomes closer to the initial destructive interference region and the gain eventually becomes negative. Since the system does not have any form of spin reservoir, saturation takes place immediately [32,33]. For extremely large $J_{C\,AC}$, $\vec{J}_{SP}$ is eventually negligible as compared to $\vec{J}_{S\,AC}$ and the gain medium is transparent.

The second mechanism we discuss is asynchronous and is based on the STT driven self-oscillations. For simplicity, we rotate the coordinate system with respect to the NM-FM bilayer as shown in Fig. 4(a). $\vec{H}_{eff}$ is set in the $\hat{z}$ direction and the oscillator is excited by injecting a DC charge current, $J_{C\,DC}$, that is converted by the SHE to a DC spin current of magnitude $J_{S\,DC}$ polarized in the $\vec{H}_{eff}$ direction. Self-oscillations in the FM are obtained at the critical $J_{C\,DC}$ value of $J_{C\,STT}$ that is given by $J_{C\,STT} = \frac{2eM_s t_{FM} H_0}{\hbar \Theta_{SHE}} \frac{\alpha + \alpha_{SP}}{1+(\alpha+\alpha_{SP})^2}$. Here as well, $\alpha_{SP}$ expresses the additional spin angular momentum losses that result from spin pumping of the FM into the NM. $\vec{J}_{S\,AC}$ consists of spins polarized along $\hat{x}$. Amplification in this case is asynchronous. The phase of the STT oscillator is locked to $\vec{J}_{S\,AC}$ and is determined during the first few cycles of the interaction. Depending on the magnitude of $J_{C\,DC}$, $\vec{J}_{S\,AC}$ is either amplified or absorbed. The gain saturation profiles are presented in Fig. 4(b). In contrast to the synchronous amplifier, now the gain profiles resemble those of the optical amplifier: the small signal gain is constant and the gain saturates at large $J_{C\,AC}$ values. The similarity to the optical amplifier stems from the analogy between the injection of incoherent carriers in the optical amplifier and the phase-less spins in the spin current amplifier. As a function of $J_{C\,DC}$, the gain profiles are antisymmetric around the critical current so that when $J_{C\,DC} < J_{C\,STT}$ absorption takes place and above $J_{C\,STT}$ amplification is reached. This behavior is summarized in Fig. 4(c).

We explain the antisymmetry using the analogy to the Bloch vector formalism [34-36]. To this end, we transform $\vec{M}$ to the density matrix elements of a TLS according to: $M_z = \rho_{11} - \rho_{22}$ and $\rho_{12} = (M_x - jM_y)/2$, where $\rho_{11}$ and $\rho_{22}$ are the occupation probabilities of the ground and excited states, respectively, and $\rho_{12}$ is the off-diagonal term of the density matrix. Under this transformation, the LLGS equation describes the dynamics of an effective TLS as follows [33,37,38] (see Supplementary Material Note 1):

$$\begin{cases} \dot{\rho}_{11} = \Lambda_1 - \gamma_1 \rho_{11} + 2[\Im(\rho_{12})\Re(V_{12}) - \Re(\rho_{12})\Im(V_{12})] \\ \dot{\rho}_{22} = \Lambda_2 - \gamma_2 \rho_{22} - 2[\Im(\rho_{12})\Re(V_{12}) - \Re(\rho_{12})\Im(V_{12})] \\ \dot{\rho}_{12} = -(j\omega + \gamma_{inh})\rho_{12} + j(\rho_{11} - \rho_{22})V_{12} \end{cases} \quad (1)$$

where $\Lambda_1$ and $\Lambda_2$ are the injection rates into the ground and excited states and $\gamma_1$ and $\gamma_2$ are the decay rates of the ground and excited states, respectively. $\omega$ is the natural resonance frequency of the TLS, $\gamma_{inh}$ is the inhomogeneous broadening of the system, and $V_{12}$ is the interaction term. $\gamma_1$ and $\gamma_2$ are related to $\alpha$ [22] and are introduced phenomenologically in the optical equations [37] in contrast to the Gilbert damping that describes a Rayleigh friction process [39]. $\Lambda_1$ and $\Lambda_2$ are relevant in the optical equations and are zero in our case. With the above transformation, $\gamma_1$ and $\gamma_2$ are given by (see Supplementary Material Note 2):

$$\begin{cases} \gamma_1(M_z) = -\gamma H_0 \dfrac{\alpha + \alpha_{SP}}{1+(\alpha+\alpha_{SP})^2}\left(\dfrac{M_s - M_z}{M_s}\right)\left[1 - \dfrac{J_{C\,DC}}{J_{C\,STT}}\right] \\ \gamma_2(M_z) = \gamma H_0 \dfrac{\alpha + \alpha_{SP}}{1+(\alpha+\alpha_{SP})^2}\left(\dfrac{M_s + M_z}{M_s}\right)\left[1 - \dfrac{J_{C\,DC}}{J_{C\,STT}}\right] \end{cases}. \quad (2)$$

Since $M_z \leq M_s$, the term $\gamma H_0 \frac{\alpha+\alpha_{SP}}{1+(\alpha+\alpha_{SP})^2}\left(\frac{M_s \pm M_z}{M_s}\right)$ is always non-negative, resulting in $\gamma_1 \propto -\left[1 - \frac{J_{C\,DC}}{J_{C\,STT}}\right]$ and $\gamma_2 \propto +\left[1 - \frac{J_{C\,DC}}{J_{C\,STT}}\right]$, so that above the critical current $\gamma_1$ and $\gamma_2$ change signs. Amplification in the TLS occurs when the population is inverted, namely when $M_z$ is negative. Below the critical current, $J_{C\,DC} < J_{C\,STT}$, $\gamma_2 > 0$ and $\gamma_1 < 0$ resulting in $M_z > 0$ and the population inversion is negative. Therefore, absorption takes place and $\phi_{J_{SP}-J_{S\,AC}}$ is such that $\vec{J}_{SP}$ and $\vec{J}_{S\,AC}$ add destructively. As compared to the synchronous amplifier where $\phi_{J_{SP}-J_{S\,AC}}$ is set by the optical pulses, here it is set by the population inversion, $\rho_{11} - \rho_{22}$. The population inversion sets the phase of $\rho_{12}$ that stands behind $\phi_{J_{SP}-J_{S\,AC}}$, as seen from the expression for $\dot{\rho}_{12}$. In a similar manner, above the critical current $\gamma_1$ and $\gamma_2$ change sign so that $\vec{M}$ amplifies $\vec{J}_{S\,AC}$. Fig. 4(d) illustrates the dependence of the gain on the population inversion. The Figure presents the temporal traces of $M_z$, the constant envelope of the injected $(\vec{J}_{S\,AC})_x$, and $(\vec{J}_{s\,out})_x$ under amplification conditions ($J_{C\,DC} > J_{C\,STT}$). It is seen that absorption takes place for positive $M_z$ and amplification for negative $M_z$. At sufficiently long times, beyond the plotted time window, the interaction reaches a steady amplification state.

The transition from absorption to amplification around $J_{C\,STT}$ is less abrupt the larger $J_{C\,AC}$ is. This is seen as well in Fig. 4(c). For small $J_{C\,AC}$ the STT oscillator dictates the dynamics. In contrast, when $J_{C\,AC}$ is large, e.g. for $J_{C\,AC} = 10^9\ A/m^2$, $J_{C\,AC}$ dominates the dynamics and in addition the gain is saturated as seen in Fig. 4(b). Therefore, the transition from absorption to amplification is moderate. At $J_{C\,DC}$ values slightly beyond $J_{C\,STT}$ of up to $\sim 1.04 \cdot J_{C\,STT}$ (marked by the dashed black frame), the behavior resembles the bias current dependent gain profile of the optical amplifier [40] presented in the inset of the Figure. Beyond $\sim 1.04 \cdot J_{C\,STT}$, the spin current gain decreases and the behavior of the spin and optical amplifiers deviate. This deviation is explained as well by the density matrix formalism. The amplified spin current is given by $(\vec{J}_{S\,out})_x = (\vec{J}_{S\,AC} + \vec{J}_{SP})_x$ in which $(\vec{J}_{SP})_x$ is proportional to $M_y \dot{M}_z - \dot{M}_y M_z$. Since $M_z$ is slowly varying as compared to $M_y$, the first term can be neglected resulting in $(\vec{J}_{SP})_x \propto \dot{M}_y M_z$ which translates to $(\vec{J}_{SP})_x \propto (\dot{\rho}_{12} - \dot{\rho}_{21})(\rho_{11} - \rho_{22})$. At the critical current $J_{C\,DC} = J_{C\,STT}$ and transparency occurs, namely, $(\rho_{11} - \rho_{22}) = 0$. On the other hand, when $J_{C\,DC} \gg J_{C\,STT}$, $\vec{M}$ becomes anti-parallel to $\vec{H}_{eff}$ because of the large anti-damping torque and the transverse magnetization components decrease, resulting in $\dot{\rho}_{12}, \dot{\rho}_{21} \to 0$. Therefore, $(\vec{J}_{SP})_x \to 0$ in both limits. It is also seen that the expression for $(\vec{J}_{SP})_x$ is anti-symmetric with respect to the transparency point at $J_{C\,STT}$.

Finally, we point out an additional difference between the spin current amplifier and the optical amplifier that is found in the DC excitation process of the TLS. In the TLS optical amplifier model, the incoherent current injection increases $\rho_{22}$ [33,41] as illustrated in Eq. (1) by the term $\Lambda_2$. In contrast, in the density matrix formalism of the spin current amplifier, the incoherent spin injection manifests through the decay terms, e.g. $\gamma_2$ in Eq. (2), by the anti-damping STT according to the LLGS equation.

**Summary**

In summary, we present two mechanisms for amplifying spin currents by stimulated spin pumping. The first mechanism is synchronous and provides control over the phase of the pumped spin current thereby revealing the inner workings of the NA amplification process. The second mechanism is asynchronous and is a closer analogue of the optical amplifier. The similarity stems from the dynamics of converting incoherent carriers/spins into coherent carriers/spins of a well-defined phase. Our findings stimulate further connections between well-established concepts from laser physics and spintronics technology. For example, a saturable spin current absorber that gives rise to a mode-locked spin current emission capable of exerting a torque that is stronger than observed to date may be considered. Likewise, Gilbert-type loss terms that benefit from a rigorous physical origin can be introduced into the optical TLS. Future work should relate to spatially distributed traveling wave propagation effects and most importantly to the experimental realization of the spin current amplifier.

Figure 1

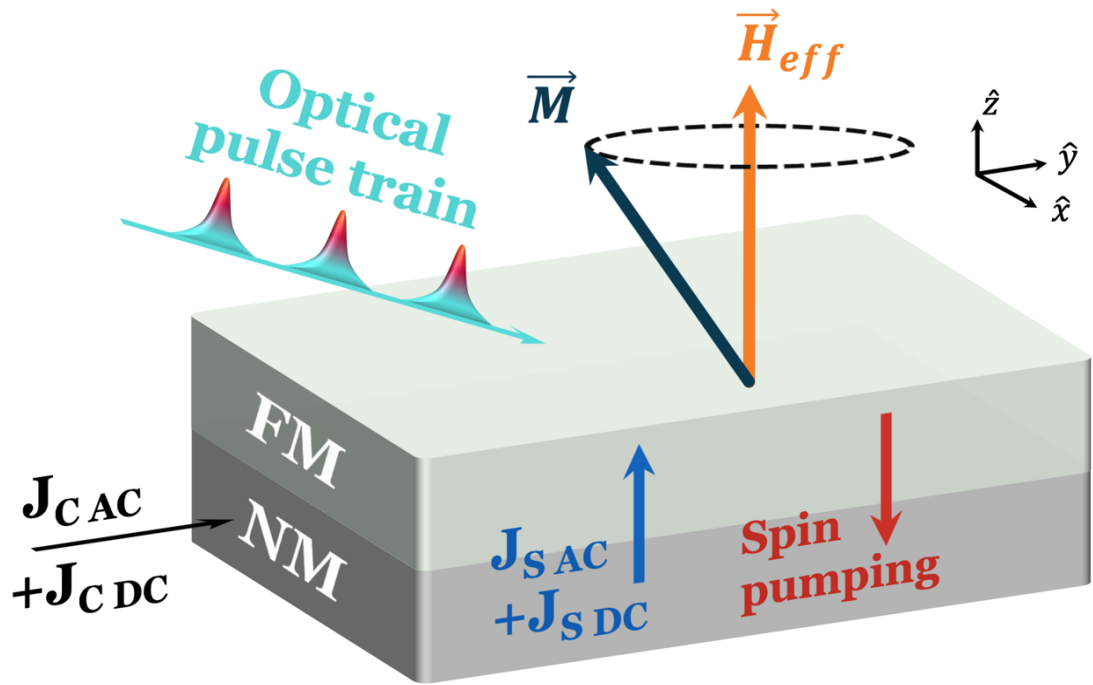

**Fig. 1.** Physical configuration of the simulation. Optical pulses and a DC charge current are applied in the synchronous and asynchronous mechanisms, respectively.

Figure 2

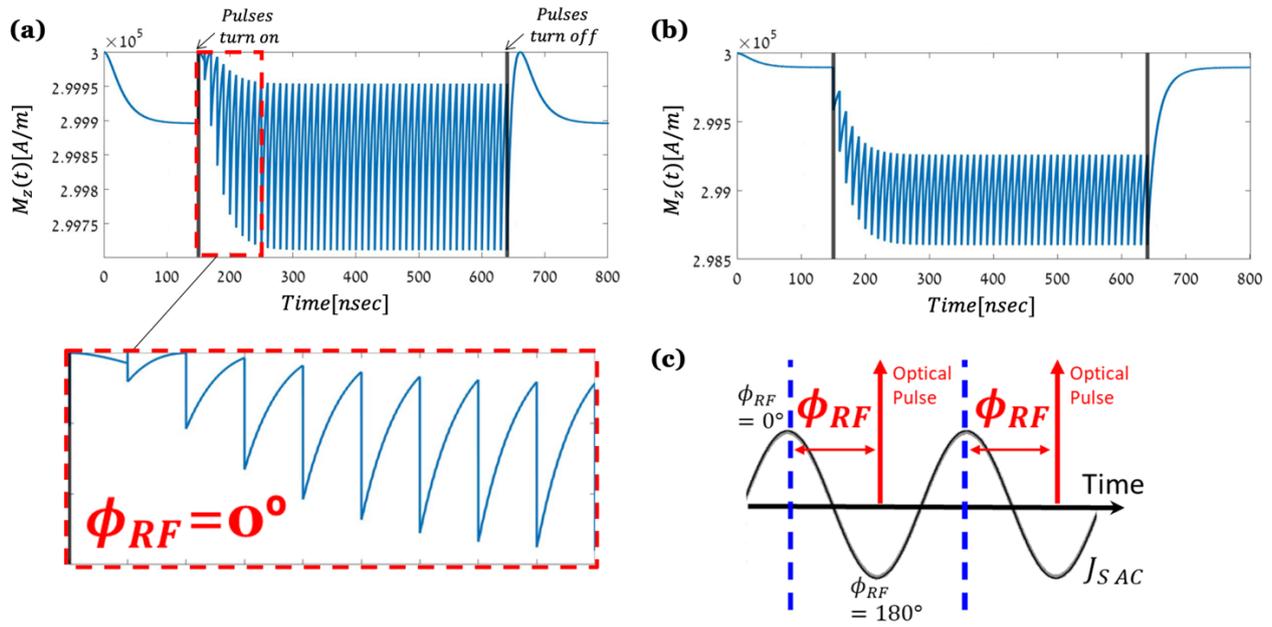

**Fig. 2. Temporal evolution of $\vec{M}$. (a) $M_z(t)$ for amplification conditions. Inset presents a close-up of t = 150 - 230 ns. (b) $M_z(t)$ for absorption conditions. In (a) and (b) the optically induced magnetic field pulses were 0.05 [T]. (c) Illustration of $\phi_{RF}$.**

**Figure 3**

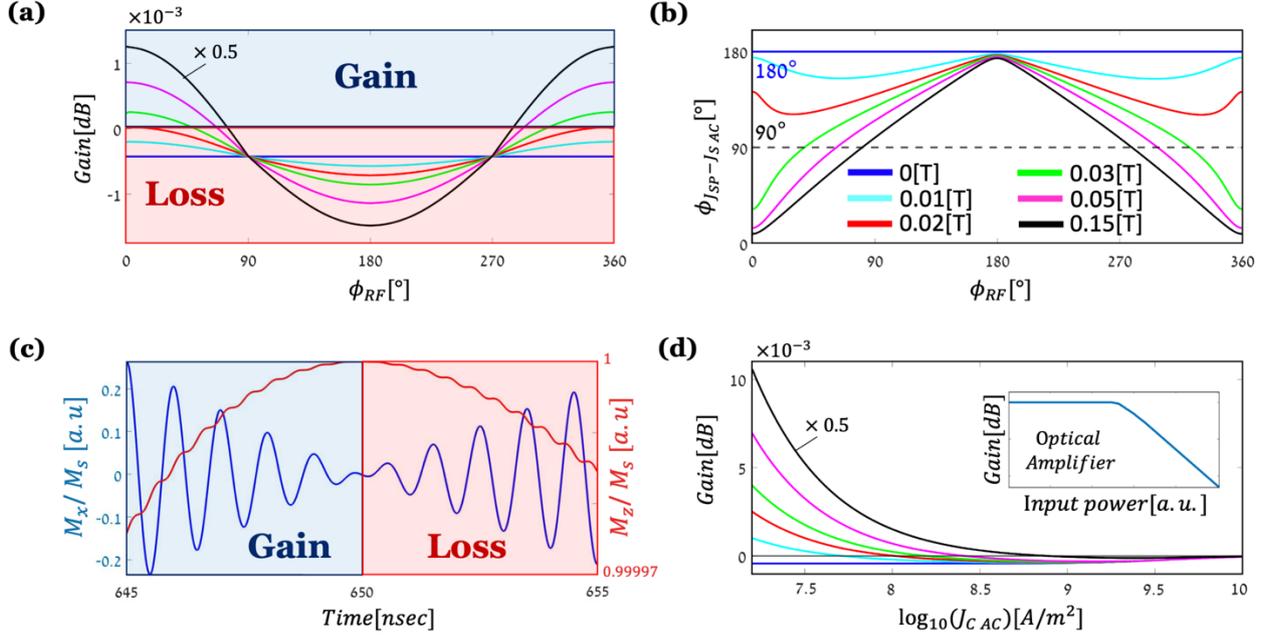

Fig. 3. Synchronous amplification. (a) $\phi_{RF}$ dependent gain. (b) $\phi_{RF}$ dependent phase between the spin pumping and the injected AC spin current. (c) Close up of times t=645 ns to t=655 ns of Fig. 2(a) illustrating the chirping of $\vec{M}$ once the pulses are turned off. (d) Gain saturation profiles for $\phi_{RF} = 0°$. Inset present the gain saturation of the optical amplifier. (a) and (b) are calculated for $J_{C\,AC} = 10^8\ A/m^2$.

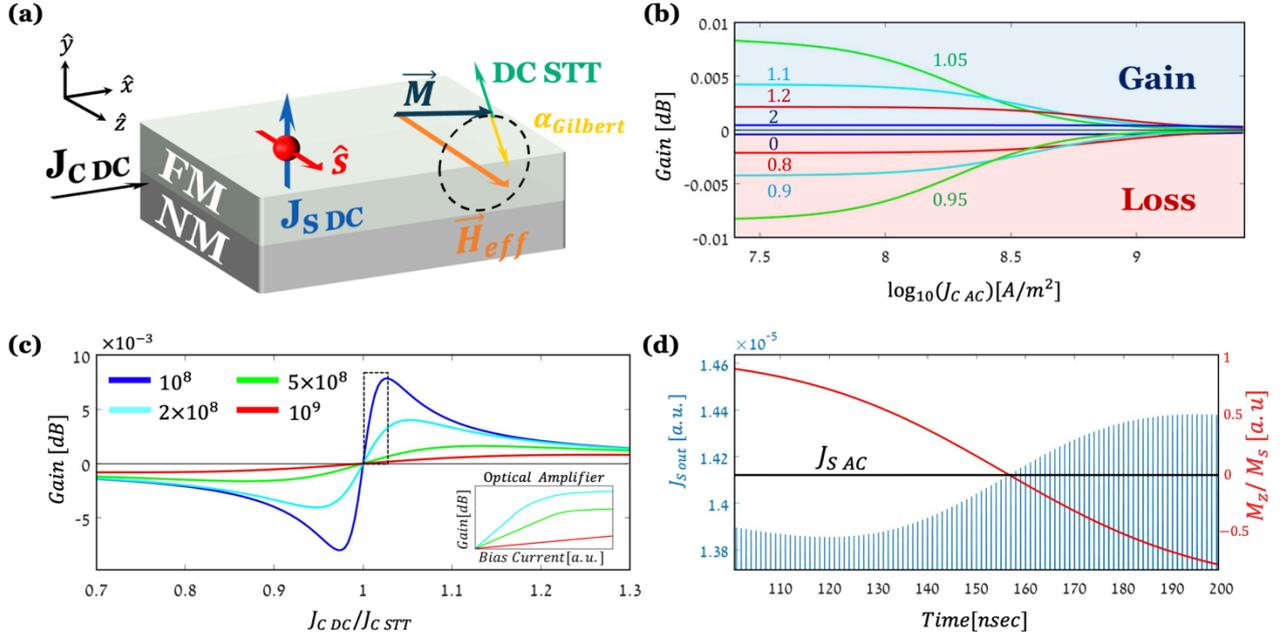

**Fig. 4. Asynchronous spin current amplification. (a)** Scheme of the STT oscillator used in the asynchronous amplification. **(b)** Gain saturation profiles for $J_{C\,DC}$ of $0 - 2 J_{C\,STT}$. Symmetric $J_{C\,DC}$ values with respect to $J_{C\,STT}$ are plotted with the same color code. **(c)** Dependence of the gain on the DC spin current for $J_{C\,AC}$ of $10^8 - 10^9\,A/m^2$. Inset presents qualitatively the bias current dependent gain saturation profile of the optical amplifier for three levels of input optical power. Red solid line in the inset corresponds to the highest input power and magenta corresponds to lowest. **(d)** Temporal traces of $J_{S\,out}$ and $M_z$ prior to reaching the quasi steady amplification state for $J_{C\,AC} = 10^8\,A/m^2$ and $J_{C\,DC} = 1.2 \cdot J_{C\,STT}$. Black solid line indicates the amplitude of the injected AC spin current, $(\vec{J}_{S\,AC})_x$. Blue solid line indicates the amplified spin current, $(\vec{J}_{S\,out})_x$. Red solid line is the normalized $M_z$ component. Transparency takes place for $M_z = 0$.